\documentclass[a4paper,aps,onecolumn,nofootinbib]{revtex4}
\RequirePackage[colorlinks,hyperindex]{hyperref}
\RequirePackage[english]{babel}
\RequirePackage[latin1]{inputenc}
\RequirePackage[T1]{fontenc}
\RequirePackage{mathrsfs}
\RequirePackage{amsmath}
\RequirePackage{amssymb}
\RequirePackage{amsbsy}
\RequirePackage{color}
\RequirePackage{bm}
\hypersetup{colorlinks=true,breaklinks=true,urlcolor=blue,linkcolor=red}
\begin{document}
\title{\bf{Polar form of Dirac fields: implementing symmetries via Lie derivative}}
\author{Luca Fabbri$^{\nabla}$\!\!\! $^{\hbar}$\footnote{luca.fabbri@edu.unige.it},
Stefano Vignolo$^{\nabla}$\!\!\! $^{\hbar}$\footnote{stefano.vignolo@unige.it},
Roberto Cianci$^{\nabla}$\!\!\! $^{\hbar}$\footnote{roberto.cianci@unige.it}}
\affiliation{$^{\nabla}$DIME, Universit\`{a} di Genova, Via all'Opera Pia 15, 16145 Genova, ITALY\\
$^{\hbar}$INFN, Sezione di Genova, Via Dodecaneso 33, 16146 Genova, ITALY}
\date{\today}
\begin{abstract}
We consider the Lie derivative along Killing vector fields of the Dirac relativistic spinors: by using the polar decomposition we acquire the mean to study the implementation of symmetries on Dirac fields. Specifically, we will become able to examine under what conditions it is equivalent to impose a symmetry upon a spinor or only upon its observables. For one physical application, we discuss the role of the above analysis for the specific spherical symmetry, obtaining some no-go theorem regarding spinors and discussing the generality of our approach.
\end{abstract}
\maketitle
\section{Introduction}
In differential geometry, the Lie derivative is a fundamental tool to implement a given symmetry upon an assigned tensor field. In the case of spherical symmetry, for instance, it is by means of such a differential operator that one is able to prove that scalars can depend only on the radial coordinate or that vectors cannot have any angular component. While the Lie derivative is well-defined and it applies on tensors of whatever rank, its extension to spinors seems to be affected by some ambiguity. The Lie derivative of spinor fields was introduced by Lichnerowicz in the 1960s \cite{Lich}, and systematized by one of his students shortly later \cite{Kos}. A treatment that gives a more geometrical meaning can be found in \cite{Jhan}. However, in the 1980s, Penrose and Rindler gave a different definition \cite{P&R}. A comparison between the two approaches and a discussion of the differential geometry that lies at their bases are in \cite{gm1, gm2} and \cite{h}. The feature that is common to all treatments, however, is that when the Lie derivative is taken along the Killing vector fields, that is along the generators of one-parameter group of local diffeomorphisms that preserve the metric of the manifold, all above definitions seem to converge to one unambiguous form. This form is the one that we are going to employ.

In our work, the Lie derivative of spinor fields will be investigated by means of the polar decomposition. The polar decomposition is the process in terms of which a complex function can be written as the product of one module times one phase. Because one particular complex function is the wave function, such polar decomposition allows us to write quantum mechanics in a form that contains only objects like density and momentum in what is called hydrodynamic form. In presence of spin and even more for relativistic processes, the wave function becomes a doublet (allowing both helicities) or a quadruplet (allowing both helicities and chiralities), and therefore the polar decomposition would have to be done in terms of several modules and phases. However, in a spinor field, the $4$ different complex spinor components mix between each other under rotations and boosts, with problems for manifest covariance. A way to write the polar decomposition of spinors in the relativistic case that is also manifestly covariant is possible \cite{jl1, jl2}. Indeed, introducing suitable quantities called tensorial connections, one can have the polar decomposition of the Dirac spinor fields extended to include differential structures \cite{Fabbri:2021mfc}, and therefore the dynamics \cite{Fabbri:2023yhl}. Having obtained the covariant derivative of Dirac spinors in polar form one can then write their Lie derivative in polar form.

The polar formalism, then, will allow us to obtain a decomposition of the Lie derivative of the spinor in terms of the Lie derivatives of its spinor bi-linears. As a consequence, we will be able to see whether the implementation of symmetries upon a spinor or upon its observables are equivalent or not.

As an application to a physical situation of large interest, we will focus on the example of stationary and spherical space-times. These back-grounds have been considered in presence of spinor fields in order to obtain exact solutions for compact objects or in cosmology \cite{Saha:2018ufp, Bronnikov:2019nqa, fss, bs-k, fss1, fss2, fss3, Herdeiro:2017fhv, Herdeiro:2019mbz, Dzhunushaliev:2019kiy, Blazquez-Salcedo:2019qrz, H, KT, Cotaescu:2007xv, Kanno:2016qcc}. So far there does not seem to be a general consensus, since some solutions may be found but they actually do not correspond to a full spherical symmetry of the spinor or on the other hand some no-go theorems are proven but they actually involve specific choices limiting their generality. In the following we will see that stationary spherical symmetry cannot be imposed onto spinor fields via Lie derivative, then no solution can be found. We will discuss how general our results in fact are as they do not involve the choice of any tetradic structure and they remain valid for whatever external interaction.

The paper is organized as follows: in section \ref{SecLie} we will recall the definition and main results about the Lie derivative of spinor fields; in section \ref{SecPol} we will present a summary of the polar form of spinor fields; in section \ref{SecLiePol} we will study the polar form of the Lie derivative finding the condition under which a symmetry implemented upon the spinor bi-linears of a spinor is equivalent to requiring that symmetry implemented upon the spinor as a whole; in section \ref{SecSpher} we apply the above results to the case of spherical symmetry, discussing a possible no-go theorem.
\section{Lie Derivative of Spinor Fields}\label{SecLie}
To begin, we recall some generalities on the Lie derivative of spinor fields $\psi$. As we have anticipated in the introduction, we will follow the definition of \cite{Lich, Kos} given by
\begin{eqnarray}
&L_{\xi}\psi\!=\!\xi^{\mu}\partial_{\mu}\psi
\!+\!\frac{1}{2}(L_{\xi}e_{k})^{\alpha}\boldsymbol{\sigma}_{\alpha}^{\phantom{\alpha}k}\psi
\label{Lie}
\end{eqnarray}
where $L_{\xi}e_{k}$ is the Lie derivative of the $k$-th vector $e_{k}$ of the basis of tetrads that we are using to move from coordinate (Greek) indices to world (Latin) indices. The Minkowski matrix $\eta_{ab}$ is used to move world indices up and down. Also, we used the convention for which $\boldsymbol{\sigma}^{ik}\!=\![\boldsymbol{\gamma}^{i},\boldsymbol{\gamma}^{k}]/4$ where $\boldsymbol{\gamma}^{a}$ are the Clifford matrices. Expression (\ref{Lie}) is writeable as
\begin{eqnarray}\label{spinor_Lie_derivative}
&L_{\xi}\psi\!=\!\xi^{\mu}\nabla_{\mu}\psi
\!+\!\frac{1}{4}(\partial \xi)_{\alpha\beta}\boldsymbol{\sigma}^{\alpha\beta}\psi
\label{Liecov}
\end{eqnarray}
with $(\partial \xi)_{\alpha\nu}\!:=\!\partial_{\alpha}\xi_{\nu}\!-\!\partial_{\nu}\xi_{\alpha}$ and $\nabla_{\mu}\psi$ denotes the spinor covariant derivative performed by means of the Levi-Civita spin connection. Now, it is known that from a given spinor field one can form spinor bi-linear quantities that are real tensors (for example, with $\overline{\psi}$ denoting the adjoint spinor, $U^{\alpha}\!=\!\overline{\psi}\boldsymbol{\gamma}^{\alpha}\psi$ is a true vector). This means that the Lie derivative of spinor fields can be used to induce a corresponding Lie derivative on their spinor bi-linear tensors fields. In the case of the vector $U^{\alpha}\!=\!\overline{\psi}\boldsymbol{\gamma}^{\alpha}\psi$ we have
\begin{eqnarray}
\nonumber
&L_{\xi}(\overline{\psi}\boldsymbol{\gamma}^{\alpha}\psi)
\!=\!L_{\xi}\overline{\psi}\boldsymbol{\gamma}^{\alpha}\psi
\!+\!\overline{\psi}L_{\xi}\boldsymbol{\gamma}^{\alpha}\psi
\!+\!\overline{\psi}\boldsymbol{\gamma}^{\alpha}L_{\xi}\psi=\\
&=\left[-\frac{1}{4}(\partial \xi)_{\nu\rho}\overline{\psi}\boldsymbol{\sigma}^{\nu\rho}
\!+\!\xi^{\mu}\nabla_{\mu}\overline{\psi}\right] \boldsymbol{\gamma}^{\alpha}\psi
\!+\!\overline{\psi}L_{\xi}\boldsymbol{\gamma}^{\alpha}\psi
\!+\!\overline{\psi}\boldsymbol{\gamma}^{\alpha}\left[\xi^{\mu}\nabla_{\mu}\psi
\!+\!\frac{1}{4}(\partial \xi)_{\nu\rho}\boldsymbol{\sigma}^{\nu\rho}\psi\right]
\end{eqnarray}
and after some re-arrangement
\begin{eqnarray}
&L_{\xi}(\overline{\psi}\boldsymbol{\gamma}^{\alpha}\psi)
\!=\!\overline{\psi}L_{\xi}\boldsymbol{\gamma}^{\alpha}\psi
\!+\!\xi^{\mu}\nabla_{\mu}(\overline{\psi}\boldsymbol{\gamma}^{\alpha}\psi)
\!-\!\frac{1}{2}(\nabla^{\nu}\xi^{\alpha}\!-\!\nabla^{\alpha}\xi^{\nu})
(\overline{\psi}\boldsymbol{\gamma}_{\nu}\psi)\label{L1}
\end{eqnarray}
in which the identity $[\boldsymbol{\sigma}^{\nu\rho},\boldsymbol{\gamma}^{\alpha}]\!=\!g^{\alpha\rho}\boldsymbol{\gamma}^{\nu}\!-\!g^{\alpha\nu}\boldsymbol{\gamma}^{\rho}$ has been used and where we have written the curl explicitly. On the other hand, the Lie derivative of a vector can be expressed as $L_{\xi}U^{\alpha}\!=\!\xi^{\mu}\nabla_{\mu}U^{\alpha}\!-\!U^{\mu}\nabla_{\mu}\xi^{\alpha}$. For the $U^{\alpha}$ above
\begin{eqnarray}
&L_{\xi}(\overline{\psi}\boldsymbol{\gamma}^{\alpha}\psi)
\!=\!\xi^{\mu}\nabla_{\mu}(\overline{\psi}\boldsymbol{\gamma}^{\alpha}\psi)
\!-\!(\overline{\psi}\boldsymbol{\gamma}^{\mu}\psi)\nabla_{\mu}\xi^{\alpha}\label{L2}.
\end{eqnarray}
Comparing \eqref{L1} and \eqref{L2} we see that 
\begin{eqnarray}
&\overline{\psi}L_{\xi}\boldsymbol{\gamma}^{\alpha}\psi
\!=\!\frac{1}{2}(\overline{\psi}\boldsymbol{\gamma}^{\nu}\psi)\nabla_{\nu}\xi^{\alpha}
\!-\!\frac{1}{2}(\overline{\psi}\boldsymbol{\gamma}^{\nu}\psi)\nabla^{\alpha}\xi_{\nu}
\!-\!(\overline{\psi}\boldsymbol{\gamma}^{\mu}\psi)\nabla_{\mu}\xi^{\alpha}
\!=\!-\frac{1}{2}(\overline{\psi}\boldsymbol{\gamma}_{\nu}\psi)
(\nabla^{\alpha}\xi^{\nu}\!+\!\nabla^{\nu}\xi^{\alpha})
\end{eqnarray}
identically. As a consequence, the definition of Lie derivative of spinors and tensors and the procedure from which we form tensors from spinors are all compatible with each other if and only if
\begin{eqnarray}
&\overline{\psi}L_{\xi}\boldsymbol{\gamma}^{\alpha}\psi\!=\!-\frac{1}{2}(\overline{\psi}\boldsymbol{\gamma}_{\nu}\psi)
(\nabla^{\alpha}\xi^{\nu}\!+\!\nabla^{\nu}\xi^{\alpha})
\end{eqnarray}
which is always true if $\xi$ is a Killing vector \cite{Jhan}. Hence, we will always assume that
\begin{eqnarray}
&L_{\xi}\boldsymbol{\gamma}^{\alpha}\!=\!0
\end{eqnarray}
leading to the fact that Lie derivation and spinor bi-linear combinations will always be mutually consistent.

Given a spinor field $\psi$ and a Killing vector field $\xi$ we will say that the spinor is \emph{strongly Lie-invariant} if
\begin{eqnarray}
&L_{\xi}\psi\!=\!0
\end{eqnarray}
holds. Instead, when all spinor bi-linears are Lie invariant along $\xi$ we say that the spinor is \emph{weakly Lie-invariant}. Along a Killing vector field, strong Lie-invariance implies weak Lie-invariance. What we wish to discuss in the present work is whether strong and weak Lie-invariance can be equivalent.
\section{Spinor Fields in Polar Form}\label{SecPol}
In the previous section we have established the convention we will follow about spinors and Clifford matrices $\boldsymbol{\gamma}_{a}$ defined to be such that $\{\boldsymbol{\gamma}_{a},\boldsymbol{\gamma}_{b}\}\!=\!2\mathbb{I}\eta_{ab}$ and from which $\boldsymbol{\sigma}^{ik}\!=\![\boldsymbol{\gamma}^{i},\boldsymbol{\gamma}^{k}]/4$ are the generators of the Lorentz group. With the identity $2i\boldsymbol{\sigma}_{ab}\!=\!\varepsilon_{abcd}\boldsymbol{\pi}\boldsymbol{\sigma}^{cd}$ one can implicitly define the fifth gamma matrix, which here we will indicate with $\boldsymbol{\pi}$ (the reasons for this convention are that there is no justification for the index five since we are not here in pentadimensional spaces and that denoting it as a gamma with no index would make it impossible to distinguish it from a normal gamma in which the index has been suppressed for compactness. The choice of the letter pi for $\boldsymbol{\pi}$ stands for \emph{parity} or \emph{projector} much in the same way the choice of sigma for $\boldsymbol{\sigma}_{ab}$ stands for \emph{spin}). By exponentiation of the generators of the Lorentz group we get an element of the Lorentz group $\boldsymbol{\Lambda}$ and so we can define the element of the spin group as $\boldsymbol{S}\!=\!\boldsymbol{\Lambda}e^{iq\alpha}$ where $\alpha$ is the phase accounting for the gauge transformation of $q$ charge. The general property of such transformations is that they verify $\boldsymbol{S}\boldsymbol{\gamma}^{a}\boldsymbol{S}^{-1}\!=\!
\boldsymbol{\Lambda}\boldsymbol{\gamma}^{a}\boldsymbol{\Lambda}^{-1}\!=\!
\boldsymbol{\gamma}^{b}(\Lambda^{-1})^{a}_{b}$ where $(\Lambda^{-1})^{i}_{a}(\Lambda^{-1})^{j}_{b}\eta_{ij}\!=\!\eta_{ab}$ so that, according to the usual denominations, $(\Lambda)^{i}_{j}$ is the real Lorentz transformation, $\boldsymbol{\Lambda}$ the complex Lorentz transformation and $\boldsymbol{S}$ the spinorial transformation. The spinor field is an object that transforms under a spinorial transformation $\boldsymbol{S}$ according to $\psi\!\rightarrow\!\boldsymbol{S}\psi$ and $\overline{\psi}\!\rightarrow\!\overline{\psi}\boldsymbol{S}^{-1}$ where $\overline{\psi}\!=\!\psi^{\dagger}\boldsymbol{\gamma}^{0}$ is the adjoint procedure.

With this pair of adjoint spinors, we can construct the spinor bi-linears
\begin{eqnarray}
&\Sigma^{ab}\!=\!2\overline{\psi}\boldsymbol{\sigma}^{ab}\boldsymbol{\pi}\psi\ \ \ \ 
\ \ \ \ \ \ \ \ M^{ab}\!=\!2i\overline{\psi}\boldsymbol{\sigma}^{ab}\psi\label{tensors}\\
&S^{a}\!=\!\overline{\psi}\boldsymbol{\gamma}^{a}\boldsymbol{\pi}\psi\ \ \ \ 
\ \ \ \ \ \ \ \ U^{a}\!=\!\overline{\psi}\boldsymbol{\gamma}^{a}\psi\label{vectors}\\
&\Theta\!=\!i\overline{\psi}\boldsymbol{\pi}\psi\ \ \ \ 
\ \ \ \ \ \ \ \ \Phi\!=\!\overline{\psi}\psi\label{scalars}
\end{eqnarray}
which are all real tensors. As is clear they are not all linearly independent and in fact we have the relation
\begin{eqnarray}
&\Sigma^{ij}\!=\!-\frac{1}{2}\varepsilon^{abij}M_{ab}
\end{eqnarray}
showing that the two antisymmetric tensors $\Sigma^{ij}$ and $M_{ab}$ are the Hodge duals of one another. Additionally, one has
\begin{eqnarray}
&M_{ab}(\Phi^{2}\!+\!\Theta^{2})\!=\!\Phi U^{j}S^{k}\varepsilon_{jkab}\!+\!\Theta U_{[a}S_{b]}
\label{M}
\end{eqnarray}
showing that if $\Phi^{2}\!+\!\Theta^{2}\!\neq\!0$ then also $M_{ab}$ can be dropped in favour of the two vectors and the two scalars. Axial-vector and vector with pseudo-scalar and scalar are also not independent since
\begin{eqnarray}
&2U_{\mu}S_{\nu}\boldsymbol{\sigma}^{\mu\nu}\boldsymbol{\pi}\psi\!+\!U^{2}\psi=0\label{AUX}
\end{eqnarray}
as well as
\begin{eqnarray}
&U_{a}U^{a}\!=\!-S_{a}S^{a}\!=\!\Theta^{2}\!+\!\Phi^{2}\label{NORM}\\
&U_{a}S^{a}\!=\!0\label{ORTHOGONAL}
\end{eqnarray}
and in the case in which $\Phi^{2}\!+\!\Theta^{2}\!\neq\!0$ we can see that the axial-vector is space-like while the vector is time-like.

Throughout this paper we will systematically deal with spinor fields for which $\Phi^{2}+\Theta^{2}\!\neq\!0$ called \emph{regular} spinor fields (spinors for which $\Phi\!\equiv\!\Theta\!\equiv\!0$ are called \emph{singular} spinor fields, or flag-dipole spinors, and they are of considerable interest \cite{Fabbri:2020elt}, since they are a class that contains also Majorana and Weyl spinors). In this case, it is always possible to write any Dirac spinor in polar form which, in chiral representation, is given as
\begin{eqnarray}
&\psi\!=\!\phi\ e^{-\frac{i}{2}\beta\boldsymbol{\pi}}
\ \boldsymbol{L}^{-1}\left(\begin{tabular}{c}
$1$\\
$0$\\
$1$\\
$0$
\end{tabular}\right)
\label{spinor}
\end{eqnarray}
for a pair of functions $\phi$ and $\beta$ and for some $\boldsymbol{L}$ having the structure of a spinorial transformation (that is, $\boldsymbol{L}$ has mathematically the same structure of $\boldsymbol{S}$) \cite{jl1, jl2}. Then
\begin{eqnarray}
&\Theta\!=\!2\phi^{2}\sin{\beta}\ \ \ \ 
\ \ \ \ \ \ \ \ \Phi\!=\!2\phi^{2}\cos{\beta}
\end{eqnarray}
showing that $\phi$ and $\beta$ are a real scalar and a real pseudo-scalar, called module and chiral angle. We can also normalize
\begin{eqnarray}
&S^{a}\!=\!2\phi^{2}s^{a}\ \ \ \ 
\ \ \ \ \ \ \ \ U^{a}\!=\!2\phi^{2}u^{a}
\end{eqnarray}
where $u^{a}$ and $s^{a}$ are the normalized velocity vector and spin axial-vector. Then (\ref{AUX}) and (\ref{NORM}-\ref{ORTHOGONAL}) reduce to
\begin{eqnarray}
&u_{[\mu}s_{\nu]}\boldsymbol{\sigma}^{\mu\nu}\boldsymbol{\pi}\psi\!+\!\psi=0\label{aux}
\end{eqnarray}
and
\begin{eqnarray}
&u_{a}u^{a}\!=\!-s_{a}s^{a}\!=\!1\label{norm}\\
&u_{a}s^{a}\!=\!0\label{orthogonal}
\end{eqnarray}
showing that the velocity has only $3$ independent components, which could be identified with the $3$ components of its spatial part, whereas the spin has only $2$ independent components, which could be identified with the $2$ angles that, in the rest-frame, its spatial part forms with the third axis. As for the matrix $\boldsymbol{L}$ we can read its meaning as that of the specific transformation that takes a given spinor into its rest frame with spin aligned along the third axis. We have already said that $\boldsymbol{L}$ has mathematically the same structure of $\boldsymbol{S}$ but it is also important to specify that from a physical perspective they are very different. In fact, while $\boldsymbol{S}$ denotes the most general spinorial transformation, $\boldsymbol{L}$ is that special spinorial transformation that takes a generic spinor in its simplest rest-frame spin-eigenstate form. Metaphorically, if the spinor were to be a top spinning on a table then $\boldsymbol{S}$ would tell how to move from the fixed system of reference in which the table is at rest to the rotating system of reference in which the top is at rest while $\boldsymbol{L}$ would tell how the top is spinning. For spinors in polar form, their $4$ complex components, or $8$ real functions, are re-organized in such a way that the $2$ real scalars $\phi$ and $\beta$ remain isolated from the $6$ parameters of $\boldsymbol{L}$ that can always be transferred into the frame and which are thus the Goldstone fields of the spinor. In fact, the Goldstone fields we have here for the spinor play the same role played by the Goldstone bosons in the Standard Model for the Higgs field. The $3$ velocities and $2$ angles amount to a total of $5$ parameters in the Lorentz transformation, and since the phase adds $1$ parameter, the full spinorial transformation has a total of $6$ parameters. Or in alternative, one could count $1$ parameter for the phase plus $6$ parameters that Lorentz transformations have in general, for a total of $7$ parameters, then subtract one parameter that is redundant, due to the fact that the gauge transformation and the rotation around the third axis have the same effect on the spinor, as is clear from (\ref{spinor}). To continue the parallel with the Standard Model, recall that the $\mathrm{U(1)\!\times\!SU(2)}$ gauge group has a total of $4$ parameters, but the hypercharge and the third component of the isospin combine to form a single parameter. Finally, to conclude this parallel, remark that here the frame in which the spinor is at rest and with spin aligned along the third axis is analogous to what in the Standard Model is for the Higgs field the unitary gauge \cite{Fabbri:2021mfc}.

Now, in general, the spinorial covariant derivative is defined according to
\begin{eqnarray}
&\nabla_{\mu}\psi\!=\!\partial_{\mu}\psi
\!+\!\boldsymbol{C}_{\mu}\psi\label{spincovder}
\end{eqnarray}
in terms of the spinorial connection $\boldsymbol{C}_{\mu}$ which is itself defined by its transformation
\begin{eqnarray}
&\boldsymbol{C}_{\mu}\!\rightarrow\!\boldsymbol{S}\left(\boldsymbol{C}_{\mu}
\!-\!\boldsymbol{S}^{-1}\partial_{\mu}\boldsymbol{S}\right)\boldsymbol{S}^{-1}
\label{spinconn}
\end{eqnarray}
where $\boldsymbol{S}$ is the spinorial transformation. This spinorial connection can be decomposed according to
\begin{eqnarray}
&\boldsymbol{C}_{\mu}\!=\!\frac{1}{2}C^{ab}_{\phantom{ab}\mu}\boldsymbol{\sigma}_{ab}
\!+\!iqA_{\mu}\boldsymbol{\mathbb{I}}\label{spinorialconnection}
\end{eqnarray}
where $C^{ab}_{\phantom{ab}\mu}$ is the spin connection of the space-time and $A_{\mu}$ is the gauge potential. Because in general
\begin{eqnarray}
&\boldsymbol{L}^{-1}\partial_{\mu}\boldsymbol{L}\!=\!iq\partial_{\mu}\zeta\mathbb{I}
\!+\!\frac{1}{2}\partial_{\mu}\zeta_{ij}\boldsymbol{\sigma}^{ij}\label{spintrans}
\end{eqnarray}
for some $\zeta$ and $\zeta_{ij}$ that are precisely the Goldstone fields of the spinor. Then we can define the quantities
\begin{eqnarray}
&P_{\mu}\!:=\!q(\partial_{\mu}\zeta\!-\!A_{\mu})\label{P}\\
&F_{ij\mu}\!:=\!\partial_{\mu}\zeta_{ij}\!-\!C_{ij\mu}\label{R}
\end{eqnarray}
which are proven to be real tensors. From (\ref{spinor}) and (\ref{P}-\ref{R}) we get
\begin{eqnarray}
&\nabla_{\mu}\psi\!=\!(-\frac{i}{2}\nabla_{\mu}\beta\boldsymbol{\pi}
\!+\!\nabla_{\mu}\ln{\phi}\mathbb{I}
\!-\!iP_{\mu}\mathbb{I}\!-\!\frac{1}{2}F_{ij\mu}\boldsymbol{\sigma}^{ij})\psi
\label{decspinder}
\end{eqnarray}
as the polar form of the covariant derivative. By taking this form and contracting on the left with $\overline{\psi}\boldsymbol{\gamma}^{k}$ we get
\begin{eqnarray}
&\overline{\psi}\boldsymbol{\gamma}^{k}\nabla_{\mu}\psi
\!=\!U^{k}\nabla_{\mu}\ln{\phi}\!-\!\frac{i}{2}\nabla_{\mu}\beta S^{k}\!-\!iP_{\mu}U^{k}
\!+\!\frac{i}{4}F_{ij\mu}\varepsilon^{kijq}S_{q}-\frac{1}{2}F_{ij\mu}\eta^{ki}U^{j}
\end{eqnarray}
whose real part is simply
\begin{eqnarray}
&\nabla_{\mu}U_{k}\!=\!2U_{k}\nabla_{\mu}\ln{\phi}+U^{j}F_{jk\mu}.
\end{eqnarray}
This can be written as
\begin{eqnarray}
&\nabla_{\mu}(2\phi^{2}u_{k})\!=\!2\phi^{2}u_{k}\nabla_{\mu}\ln{\phi^{2}}
+2\phi^{2}u^{j}F_{jk\mu}
\end{eqnarray}
and so
\begin{eqnarray}
&2\nabla_{\mu}\phi^{2}u_{k}+2\phi^{2}\nabla_{\mu}u_{k}
\!=\!2\phi^{2}u_{k}\nabla_{\mu}\ln{\phi^{2}}+2\phi^{2}u^{j}F_{jk\mu}
\end{eqnarray}
where the derivatives of the module eventually cancel. Therefore
\begin{eqnarray}
&\nabla_{\mu}u_{k}\!=\!u^{j}F_{jk\mu}
\end{eqnarray}
as a general identity. The same is true for the axial-vector. In conclusion, we have
\begin{eqnarray}
&\nabla_{\mu}s_{i}\!=\!F_{ji\mu}s^{j}\ \ \ \
\ \ \ \ \ \ \ \ \nabla_{\mu}u_{i}\!=\!F_{ji\mu}u^{j}\label{ds-du}
\end{eqnarray}
which are valid as general identities. These last two identities imply in turn the validity of the relation
\begin{eqnarray}
&F_{ab\mu}\!\equiv\!u_{a}\nabla_{\mu}u_{b}\!-\!u_{b}\nabla_{\mu}u_{a}
\!+\!s_{b}\nabla_{\mu}s_{a}\!-\!s_{a}\nabla_{\mu}s_{b}
\!+\!(u_{a}s_{b}\!-\!u_{b}s_{a})\nabla_{\mu}u_{k}s^{k}
\!+\!\frac{1}{2}F_{ij\mu}\varepsilon^{ijcd}\varepsilon_{abpq}s_{c}u_{d}s^{p}u^{q}
\end{eqnarray}
which can be written as
\begin{eqnarray}
&F_{ab\mu}\!=\!u_{a}\nabla_{\mu}u_{b}\!-\!u_{b}\nabla_{\mu}u_{a}
\!+\!s_{b}\nabla_{\mu}s_{a}\!-\!s_{a}\nabla_{\mu}s_{b}
\!+\!(u_{a}s_{b}\!-\!u_{b}s_{a})\nabla_{\mu}u_{k}s^{k}
\!+\!2\varepsilon_{abij}u^{i}s^{j}V_{\mu}\label{Rfull}
\end{eqnarray}
for some vector
\begin{eqnarray}
&V_{\mu}\!:=\!\frac{1}{4}F_{ij\mu}\varepsilon^{ijcd}u_{c}s_{d}\label{V}
\end{eqnarray}
that cannot be specified in terms of covariant derivatives of velocity and spin. Finally, we can write \eqref{decspinder} with \eqref{Rfull} as
\begin{eqnarray}
&\!\!\!\!\nabla_{\mu}\psi
\!=\![-\frac{i}{2}\nabla_{\mu}\beta\boldsymbol{\pi}
\!+\!\nabla_{\mu}\ln{\phi}\mathbb{I}
\!-\!i(P_{\mu}\!-\!V_{\mu})\mathbb{I}\!-\!(u_{a}\nabla_{\mu}u_{b}
\!+\!s_{b}\nabla_{\mu}s_{a}
\!+\!u_{a}s_{b}\nabla_{\mu}u_{k}s^{k})\boldsymbol{\sigma}^{ab}]\psi
\label{derpolR}
\end{eqnarray}
having used $2i\boldsymbol{\sigma}_{ab}\!=\!\varepsilon_{abcd}\boldsymbol{\pi}\boldsymbol{\sigma}^{cd}$ and (\ref{aux}). After that the Goldstone fields are transferred into the frame, they combine with spin connection and gauge potential to become the longitudinal components of the $P_{\mu}$ and $F_{ij\mu}$ tensors. The $P_{\mu}$ and $F_{ij\mu}$ objects, therefore, have the same information content of spin connection and gauge potential while being real tensors, and it is for this reason that they are called space-time and gauge tensorial connections. In this, the tensorial connections we have here are the geometric and electrodynamic analog of the weak bosons $W_{\nu}^{\pm}$ and $Z_{\nu}$ of the Standard Model. Velocity and spin can only determine $5$ parameters of $\boldsymbol{L}$ and so their derivatives can only determine $20$ components of $F_{ij\mu}$ with the $4$ missing components corresponding to rotations around the third axis encoded by the $F_{12\mu}$ component. Because for rest-frame spin-eigenstate spinors it is $u^{0}\!=\!1$ and $s^{3}\!=\!1$ it follows that $2V_{\mu}\!=\!F_{12\mu}$ and thus the missing components are encoded in the $V_{\mu}$ vector. Alternatively, one can count the $24$ components of the space-time tensorial connection plus the $4$ components of the gauge tensorial connection, totalling $28$ components, and subtract the four components that are redundant, due to the fact that $V_{\mu}$ and $P_{\mu}$ act identically on the spinor field, as is clear from the structure of (\ref{derpolR}) \cite{Fabbri:2021mfc}.

It is also important to notice that the relations
\begin{eqnarray}
&R_{\alpha\rho\mu\nu}\!=\!-(\nabla_{\mu}F_{\alpha\rho\nu}\!-\!\nabla_{\nu}F_{\alpha\rho\mu}
\!+\!F_{\alpha\kappa\mu}F_{\eta\rho\nu}g^{\kappa\eta}
\!-\!F_{\alpha\kappa\nu}F_{\eta\rho\mu}g^{\kappa\eta})\label{Riemann}\\
&qF_{\mu\nu}\!=\!-(\nabla_{\mu}P_{\nu}\!-\!\nabla_{\nu}P_{\mu})\label{Faraday}
\end{eqnarray}
are valid as geometric identities, which means that the $F_{ab\mu}$ and $P_{\mu}$ tensors can be respectively seen as gauge-invariant and covariant potentials of the Riemann curvature and the Maxwell strength.

Due to the importance of the tensorial connection, and specifically the space-time tensorial connection $F_{ab\mu}$ we now give further comments about its geometrical interpretation. To this end, let us suppose to perform the pointwise tetrad transformation given by
\begin{equation}
\hat{e}^{i}\!=\!\Lambda^{i}_{\phantom{i}a}e^{a}\ \ \ \ \ \ \ \ 
\ \ \ \ \ \ \ \ \hat{e}_{a}\!=\!\Lambda_{a}^{\phantom{a}i}e_{i}
\label{trasformazione_tetrade}
\end{equation}
where $\Lambda_{a}^{\;\;i}$ denotes a suitable real Lorentz transformation and $\Lambda^i_{\;\;a}\!=\!(\Lambda^{-1})_{a}^{\;\;i}$ is its inverse. Real Lorentz transformations are tied to the corresponding complex Lorentz transformation (or spinorial transformation) $\boldsymbol{L}$ by the requirement
\begin{equation}
\label{spinor_representation}
\boldsymbol{L}\boldsymbol{\gamma}^{j}\boldsymbol{L}^{-1}\Lambda^i_{\;\;j}=\boldsymbol{\gamma}^{i}
\end{equation}
where both real and complex (or spinorial) transformations have the same parameters, which are in general functions of the space-time coordinates. Because of the change of trivialization induced by the real Lorentz transformation, the spinor field $\psi$ undergoes the corresponding spinorial transformation
\begin{equation}
\label{trasformazione_spinore}
\hat{\psi}\!=\!\boldsymbol{L}(\Lambda^{-1})\psi
\end{equation}
by construction. In the same way, a spin connection $\Omega^{ij}_{\;\;\;\mu}$ on the space-time transforms according to
\begin{equation}
\label{trasformazione_spin_connection}
\hat{\Omega}^{ij}_{\;\;\;\mu} = \Omega^{hk}_{\;\;\;\mu}\Lambda^i_{\;\;h}\Lambda^j_{\;\;k} - \Lambda^{js}\frac{\partial}{\partial x^\mu}\Lambda^i_{\;\;s}
\end{equation}
in which we have denoted $\Lambda^{js}=\Lambda^{j}_{\;\;t}\eta^{ts}$ for brevity. In particular, we consider the flat spin connection whose coefficients $\Omega^{ij}_{\;\;\;\mu}$ are zero in the frame where the spinor is at rest and with spin aligned along the third axis, which in the parallel to the Standard Model we will call the \emph{unitary frame}. Correspondingly, we will call such spin connection the \emph{Goldstone connection}. After the transformation \eqref{trasformazione_tetrade}, in the new tetrad $\hat{e}_i$ the Goldstone connection is
\begin{equation}
\label{new_coefficients}
\hat{\Omega}^{ij}_{\;\;\;\mu}\!=\!-\Lambda^{js}\frac{\partial}{\partial x^\mu}\Lambda^i_{\;\;s}
\end{equation}
and identity \eqref{spintrans} can be written as
\begin{eqnarray}
&\frac{1}{2}\Lambda^{js}\partial_{\mu}\Lambda^i_{\;\;s}\boldsymbol{\sigma}_{ij}\boldsymbol{L}
\!=\!\partial_{\mu}\boldsymbol{L}\label{spintransprime}
\end{eqnarray}
where $\boldsymbol{L}=\boldsymbol{L}(\Lambda^{-1})$ is the matrix appearing in equation (\ref{trasformazione_spinore}). Indeed one has
\begin{eqnarray}
\nonumber
&\frac{1}{2}\Lambda^{js}\partial_{\mu}\Lambda^i_{\;\;s}\boldsymbol{\sigma}_{ij}\boldsymbol{L}
\!=\!\frac{1}{8}\left[\partial_{\mu}(\Lambda^i_{\;\;s}\boldsymbol{\gamma}_{i}),
\Lambda^{js}\boldsymbol{\gamma}_{j}\right]\boldsymbol{L}
\!=\!\frac{1}{8}\left[\partial_{\mu}(\boldsymbol{L}\boldsymbol{\gamma}_{s}\boldsymbol{L}^{-1}),
\eta^{st}\boldsymbol{L}\boldsymbol{\gamma}_{t}\boldsymbol{L}^{-1}\right]\boldsymbol{L}=\\
&=\frac{1}{4}\partial_{\mu}(\boldsymbol{L}\boldsymbol{\gamma}_{s}\boldsymbol{L}^{-1})
(\boldsymbol{L}\boldsymbol{\gamma}^{s}\boldsymbol{L}^{-1})\boldsymbol{L}
\!=\!\partial_{\mu}\boldsymbol{L}\!+\!\frac{1}{4}\boldsymbol{L}\boldsymbol{\gamma}_{s}
\partial_{\mu}\boldsymbol{L}^{-1}\boldsymbol{L}\boldsymbol{\gamma}^{s}
\!=\!\partial_{\mu}\boldsymbol{L}
\end{eqnarray}
in view of identities $\boldsymbol{\gamma}_s\left[\boldsymbol{\gamma}_i,\boldsymbol{\gamma}_j\right]\boldsymbol{\gamma}^s\!=\!0$ holding in $4$ dimensions. From \eqref{spintransprime} we derive the relation
\begin{eqnarray}
\label{relazione2}
\Lambda^{js}\frac{\partial}{\partial x^\mu}\Lambda^i_{\;\;s}\frac{1}{2}\boldsymbol{\sigma}_{ij}
\!=\!\frac{\partial \boldsymbol{L}}{\partial x^\mu}\boldsymbol{L}^{-1}
\!=\!-\frac{1}{2}\frac{\partial \zeta^{ij}}{\partial x^\mu}\boldsymbol{\sigma}_{ij}
\end{eqnarray}
and because of the linear independence of the $\boldsymbol{\sigma}_{ij}$ matrices we conclude that
\begin{eqnarray}
\label{identityaux}
\frac{\partial \zeta^{ij}}{\partial x^\mu} = - \Lambda^{js}\frac{\partial}{\partial x^\mu}\Lambda^i_{\;\;s}
\end{eqnarray}
as a general relation. The space-time tensorial connection is defined as
\begin{eqnarray}
\label{definizioneR}
F^{ij}_{\;\;\;\mu} = \frac{\partial \zeta^{ij}}{\partial x^\mu}-C^{ij}_{\;\;\;\mu}
\end{eqnarray}
where $C^{ij}_{\;\;\;\mu}$ denotes the Levi-Civita spin connection. Making use of equations \eqref{new_coefficients} and \eqref{identityaux} we can re-write \eqref{definizioneR} as
\begin{eqnarray}
\label{definizioneR2}
F^{ij}_{\;\;\;\mu} = \hat{\Omega}^{ij}_{\;\;\;\mu}\!-\!C^{ij}_{\;\;\;\mu}
\end{eqnarray}
which represents the difference between the Goldstone connection and the Levi-Civita spin connection, clarifying the tensor nature of the quantity $F^{ij}_{\;\;\;\mu}$ and its meaning. More precisely, it is evident that the tensor $F_{ab\mu}$ identifies with the contorsion tensor of the Goldstone connection. In such a circumstance, the conditions that the vectors $u^i$ and $s^i$ must satisfy (\ref{ds-du}) is the same as requiring that both vectors $u^i$ and $s^i$ verify
\begin{eqnarray}
&\hat{\nabla}_\mu u^i = 0\ \ \ \ \ \ \ \ \hat{\nabla}_\mu s^i = 0
\end{eqnarray}
that is they are constant with respect to the Goldstone covariant derivative.

Having constructed all kinematic objects we need, we are able to outline the dynamics of Dirac fields in polar form.

It is seen that the Dirac equation
\begin{eqnarray}
&i\boldsymbol{\gamma}^{\mu}\nabla_{\mu}\psi\!-\!m\psi\!=\!0
\end{eqnarray}
can be written equivalently in polar form as \cite{Fabbri:2021mfc}
\begin{eqnarray}
&\nabla_{\mu}\ln{\phi^{2}}\!+\!F_{\mu\nu}^{\phantom{\mu\nu}\nu}
\!-\!2P^{\rho}u^{\nu}s^{\alpha}\varepsilon_{\mu\rho\nu\alpha}\!+\!2ms_{\mu}\sin{\beta}\!=\!0
\label{D1}\\
&\nabla_{\mu}\beta\!+\!\frac{1}{2}\varepsilon_{\mu\alpha\nu\iota}F^{\alpha\nu\iota}
\!-\!2P^{\iota}u_{[\iota}s_{\mu]}\!+\!2ms_{\mu}\cos{\beta}\!=\!0
\label{D2}.
\end{eqnarray}
Making use of expression \eqref{Rfull}, (\ref{D1}-\ref{D2}) become
\begin{eqnarray}
&\nabla_{\mu}\ln{\phi^{2}}\!+\!u_{\mu}\nabla_{i}u^{i}\!-\!s_{\mu}\nabla_{i}s^{i}
\!-\!u^{\nu}\nabla_{\nu}u_{\mu}\!+\!s^{\nu}\nabla_{\nu}s_{\mu}
\!+\!(u_{\mu}s^{\nu}\!-\!u^{\nu}s_{\mu})\nabla_{\nu}u_{i}s^{i}
\!-\!2(P^{\rho}\!-\!V^{\rho})u^{\nu}s^{\alpha}\varepsilon_{\mu\rho\nu\alpha}
\!+\!2ms_{\mu}\sin{\beta}\!=\!0\\
&\nabla_{\mu}\beta\!-\!\varepsilon_{\mu\alpha\nu\beta}u^{\alpha}\nabla^{\nu}u^{\beta}
\!+\!\varepsilon_{\mu\beta\nu\alpha}s^{\beta}\nabla^{\nu}s^{\alpha}
\!+\!\varepsilon_{\mu\alpha\beta\nu}u^{\alpha}s^{\beta}\nabla^{\nu}u^{i}s_{i}
\!-\!2(P^{\nu}\!-\!V^{\nu})u_{[\nu}s_{\mu]}
\!+\!2ms_{\mu}\cos{\beta}\!=\!0.
\end{eqnarray}

For simplicity, let us introduce the notations
\begin{eqnarray}
&2Z_{\mu}\!:=\!\nabla_{\mu}\ln{\phi^{2}}\!+\!u_{\mu}\nabla_{i}u^{i}\!-\!s_{\mu}\nabla_{i}s^{i}
\!-\!u^{\nu}\nabla_{\nu}u_{\mu}\!+\!s^{\nu}\nabla_{\nu}s_{\mu}
\!+\!(u_{\mu}s^{\nu}\!-\!u^{\nu}s_{\mu})\nabla_{\nu}u_{i}s^{i}\label{z}\\
&2Y_{\mu}\!:=\!\nabla_{\mu}\beta\!-\!\varepsilon_{\mu\alpha\nu\beta}u^{\alpha}\nabla^{\nu}u^{\beta}
\!+\!\varepsilon_{\mu\beta\nu\alpha}s^{\beta}\nabla^{\nu}s^{\alpha}
\!+\!\varepsilon_{\mu\alpha\beta\nu}u^{\alpha}s^{\beta}\nabla^{\nu}u^{i}s_{i}\label{y}
\end{eqnarray}
so that the Dirac equations in polar form can be expressed in the more compact form
\begin{eqnarray}
&Z_{\mu}\!-\!(P^{\rho}\!-\!V^{\rho})u^{\nu}s^{\alpha}\varepsilon_{\mu\rho\nu\alpha}
\!+\!ms_{\mu}\sin{\beta}\!=\!0\label{SM}\\
&Y_{\mu}\!-\!(P^{\nu}\!-\!V^{\nu})u_{[\nu}s_{\mu]}
\!+\!ms_{\mu}\cos{\beta}\!=\!0\label{CA}.
\end{eqnarray}
Taking \eqref{SM} multiplied by $\varepsilon^{\mu\eta\pi\tau}u_{\pi}s_{\tau}$ plus \eqref{CA} multiplied by $u^{[\mu}s^{\eta]}$ gives
\begin{eqnarray}
&P^{\eta}\!-\!V^{\eta}\!=\!m\cos{\beta}u^{\eta}\!+\!Y_{\mu}u^{[\mu}s^{\eta]}
\!+\!Z_{\mu}u_{\pi}s_{\tau}\varepsilon^{\mu\pi\tau\eta}\label{momentum}
\end{eqnarray}
in which the vector $P^{\eta}\!-\!V^{\eta}$ has been made explicit. Because $\nabla_{i}S^{i}\!=\!4m\phi^{2}\sin{\beta}$ \cite{Fabbri:2023yhl}, the classical limit $S^{i}\!\rightarrow\!0$ implies also $\beta\!\rightarrow\!0$ so that \eqref{momentum} reduces to $P^{\eta}\!=\!mu^{\eta}$ showing that $P_{\eta}$ is the momentum of the particle. The general expression (\ref{momentum}) is therefore an extension of the momentum so to include the spin. Hence, the vector $P_{\mu}\!-\!V_{\mu}$ can be interpreted as the most general momentum after the redundancy between gauge and rotation around the third axis is removed.

The Dirac field has energy tensor given by
\begin{eqnarray}
&T^{\rho\sigma}
\!=\!\frac{i}{4}(\overline{\psi}\boldsymbol{\gamma}^{\rho}\nabla^{\sigma}\psi
\!-\!\nabla^{\sigma}\overline{\psi}\boldsymbol{\gamma}^{\rho}\psi
\!+\!\overline{\psi}\boldsymbol{\gamma}^{\sigma}\nabla^{\rho}\psi
\!-\!\nabla^{\rho}\overline{\psi}\boldsymbol{\gamma}^{\sigma}\psi)
\end{eqnarray}
which in polar variables becomes
\begin{eqnarray}
&T^{\rho\sigma}\!=\!\phi^{2}(P^{\rho}u^{\sigma}\!+\!P^{\sigma}u^{\rho}
\!+\!s^{\sigma}\nabla^{\rho}\beta/2
\!+\!s^{\rho}\nabla^{\sigma}\beta/2
\!-\!\frac{1}{4}F_{\alpha\nu}^{\phantom{\alpha\nu}\sigma}s_{\kappa}
\varepsilon^{\rho\alpha\nu\kappa}
\!-\!\frac{1}{4}F_{\alpha\nu}^{\phantom{\alpha\nu}\rho}s_{\kappa}
\varepsilon^{\sigma\alpha\nu\kappa})
\end{eqnarray}
where \eqref{decspinder} was used. By plugging now \eqref{Rfull} we obtain
\begin{eqnarray}
\nonumber
&T^{\rho\sigma}\!=\!\phi^{2}[(P^{\rho}\!-\!V^{\rho})u^{\sigma}
\!+\!(P^{\sigma}\!-\!V^{\sigma})u^{\rho}
\!+\!s^{\sigma}\nabla^{\rho}\beta/2
\!+\!s^{\rho}\nabla^{\sigma}\beta/2-\\
&-\frac{1}{2}s_{\kappa}u_{\alpha}\nabla^{\sigma}u_{\nu}\varepsilon^{\rho\nu\kappa\alpha}
\!-\!\frac{1}{2}s_{\kappa}u_{\alpha}\nabla^{\rho}u_{\nu}\varepsilon^{\sigma\nu\kappa\alpha}].
\label{energy}
\end{eqnarray}

As is clear from the treatment of the dynamics in polar form, the vector $P_{\mu}\!-\!V_{\mu}$ has a very special role and such a special role will become even more prominent when the Lie derivative of spinor fields will be considered.
\section{Lie Derivative in Polar Form}\label{SecLiePol}
In section \ref{SecLie} and \ref{SecPol} we have respectively defined the Lie derivative of spinor fields \eqref{Liecov} and the spinor field covariant derivative \eqref{derpolR} and it is our goal now to combine the two. Substituting \eqref{derpolR} into \eqref{Liecov} one gets
\begin{eqnarray}\label{final_equation}
\nonumber
&L_{\xi}\psi\!=\!-i\boldsymbol{\pi}\psi L_{\xi}\beta/2\!+\!\psi L_{\xi}\ln{\phi}
\!-\!\boldsymbol{\sigma}_{\alpha\rho}\psi u^{\alpha}(L_{\xi}u)^{\rho}
\!-\!\boldsymbol{\sigma}_{\alpha\rho}\psi s^{\rho}(L_{\xi}s)^{\alpha}
\!-\!\boldsymbol{\sigma}_{\alpha\rho}\psi u^{\alpha}s^{\rho}s_{\nu}(L_{\xi}u)^{\nu}+\\
&+[\frac{1}{4}(\partial \xi)_{\mu\nu}(g^{\alpha\mu}\!-\!u^{\alpha}u^{\mu}\!+\!s^{\alpha}s^{\mu})
(g^{\rho\nu}\!-\!u^{\rho}u^{\nu}\!+\!s^{\rho}s^{\nu})\boldsymbol{\sigma}_{\alpha\rho}
\!-\!i\xi^{\mu}(P_{\mu}\!-\!V_{\mu})\mathbb{I}]\psi
\end{eqnarray}
where the Lie derivative of vectors and scalars was used. It has now become evident how employing the polar decomposition has enabled us to write the Lie derivative of the Dirac spinor in terms of the Lie derivative of the spinor bi-linears. In fact, this was made possible by expressing the covariant derivative of the spinor (appearing in eq. \eqref{spinor_Lie_derivative}) in terms of the covariant derivative of the tensor and pseudo-tensor quantities that characterize the spinor in polar form. Then using the well-known relationship linking Lie derivative, covariant derivative and torsion (null in our case), we were able to obtain the final equation \eqref{final_equation}.

Eq. \eqref{final_equation} allows us to answer the question with which we have concluded section \ref{SecLie}: that is, along Killing vector fields, the strong Lie-invariance amounts to the weak Lie-invariance plus the additional condition
\begin{eqnarray}
&[\frac{1}{4}(\partial \xi)_{\mu\nu}(g^{\alpha\mu}\!-\!u^{\alpha}u^{\mu}\!+\!s^{\alpha}s^{\mu})
(g^{\rho\nu}\!-\!u^{\rho}u^{\nu}\!+\!s^{\rho}s^{\nu})\boldsymbol{\sigma}_{\alpha\rho}
\!-\!i\xi^{\mu}(P_{\mu}\!-\!V_{\mu})\mathbb{I}]\psi\!=\!0\label{cond}
\end{eqnarray}
which, however, is not of immediate meaning, yet. To work it into a simpler form, define
\begin{eqnarray}
&\omega^{\alpha\rho}\!:=\!\frac{1}{2}(\partial \xi)_{\mu\nu}(g^{\alpha\mu}\!-\!u^{\alpha}u^{\mu}\!+\!s^{\alpha}s^{\mu})
(g^{\rho\nu}\!-\!u^{\rho}u^{\nu}\!+\!s^{\rho}s^{\nu})
\label{definition}
\end{eqnarray}
so to write \eqref{cond} in the more compact form
\begin{eqnarray}
&[\frac{1}{2}\omega^{\alpha\rho}\boldsymbol{\sigma}_{\alpha\rho}
\!-\!i\xi^{\mu}(P_{\mu}\!-\!V_{\mu})\mathbb{I}]\psi\!=\!0.
\label{initcond}
\end{eqnarray}
Multiplying on the left by $\overline{\psi}\boldsymbol{\sigma}_{ab}$ and taking the real part gives
\begin{eqnarray}
&\frac{1}{2}\varepsilon_{abij}\omega^{ij}\Theta\!-\!\omega_{ab}\Phi
\!-\!2(P_{\mu}\!-\!V_{\mu})\xi^{\mu}M_{ab}\!=\!0\label{1}
\end{eqnarray}
where $2\{\boldsymbol{\sigma}_{ab},\boldsymbol{\sigma}_{cd}\}\!=\!(\eta_{ad}\eta_{bc}\!-\!\eta_{ac}\eta_{bd})\mathbb{I}\!+\!i\varepsilon_{abcd}\boldsymbol{\pi}$ with (\ref{tensors}) and (\ref{scalars}) were used. This last expression has Hodge dual
\begin{eqnarray}
&-\omega_{ab}\Theta\!-\!\frac{1}{2}\omega^{ij}\varepsilon_{ijab}\Phi
\!-\!(P_{\mu}\!-\!V_{\mu})\xi^{\mu}M^{ij}\varepsilon_{ijab}\!=\!0\label{2}
\end{eqnarray}
so that by adding \eqref{1} multiplied by $\Phi$ to \eqref{2} multiplied by $\Theta$ one gets
\begin{eqnarray}
&-\omega_{ab}(\Theta^{2}\!+\!\Phi^{2})\!-\!2(P_{\mu}\!-\!V_{\mu})\xi^{\mu}
(M_{ab}\Phi\!+\!\frac{1}{2}M^{ij}\varepsilon_{ijab}\Theta)\!=\!0
\end{eqnarray}
and after using \eqref{M} one eventually obtains
\begin{eqnarray}
&-\omega_{ab}\!-\!2(P_{\mu}\!-\!V_{\mu})\xi^{\mu}u^{j}s^{k}\varepsilon_{jkab}\!=\!0\label{omega}
\end{eqnarray}
as a real tensorial relation that is certainly simpler than the initial (\ref{initcond}), although in the process of deriving it we may have lost information. To see that we did not, let us take \eqref{omega} written so to have $\omega_{ab}$ explicit. We can then see that
\begin{eqnarray}
&[\frac{1}{2}\omega^{\alpha\rho}\boldsymbol{\sigma}_{\alpha\rho}
\!-\!i\xi^{\mu}(P_{\mu}\!-\!V_{\mu})\mathbb{I}]\psi
\!=\!-(P_{\mu}\!-\!V_{\mu})\xi^{\mu}(u^{j}s^{k}\varepsilon_{jkab}\boldsymbol{\sigma}^{ab}
\!+\!i\mathbb{I})\psi
\end{eqnarray}
and by employing once again $2i\boldsymbol{\sigma}_{ab}\!=\!\varepsilon_{abcd}\boldsymbol{\pi}\boldsymbol{\sigma}^{cd}$ and \eqref{aux} one sees that
\begin{eqnarray}
&-(P_{\mu}\!-\!V_{\mu})\xi^{\mu}(u^{j}s^{k}\varepsilon_{jkab}\boldsymbol{\sigma}^{ab}
\!+\!i\mathbb{I})\psi
\!=\!-i(P_{\mu}\!-\!V_{\mu})\xi^{\mu}(2u^{j}s^{k}\boldsymbol{\sigma}_{jk}\boldsymbol{\pi}
\!+\!\mathbb{I})\psi\!=\!0
\end{eqnarray}
showing that \eqref{initcond} is valid. Because (\ref{initcond}) is implied as well as implies (\ref{omega}), the two are equivalent. Using in \eqref{omega} the definition \eqref{definition} we reach
\begin{eqnarray}
&\frac{1}{2}(\partial \xi)_{\mu\nu}(g^{\alpha\mu}\!-\!u^{\alpha}u^{\mu}\!+\!s^{\alpha}s^{\mu})
(g^{\rho\nu}\!-\!u^{\rho}u^{\nu}\!+\!s^{\rho}s^{\nu})\!=\!-2(P_{\mu}\!-\!V_{\mu})\xi^{\mu}u_{\eta}s_{\pi}\varepsilon^{\eta\pi\alpha\rho}
\end{eqnarray}
which is antisymmetric in its two free indices and with no projection along velocity and spin vectors. Due to its antisymmetry and the fact that it takes values in a bi-dimensional plane, it has one single independent component, and so it is with no loss of generality that we can multiply it by $u^{\sigma}s^{\tau}\varepsilon_{\sigma\tau\alpha\rho}$ to get
\begin{eqnarray}
&\frac{1}{4}(\partial \xi)_{\mu\nu}s_{\tau}u_{\sigma}\varepsilon^{\mu\nu\tau\sigma}
\!=\!2\xi^{\mu}(P_{\mu}\!-\!V_{\mu})\label{COND}
\end{eqnarray}
as a real scalar relation that is the simplest we can get and which is the equivalent of (\ref{cond}). Thus, we have proven the

\textbf{Theorem} - \emph{Let $\xi$ be a Killing vector field and $\psi$ a Dirac spinor: the validity of the relationship \eqref{COND} is a sufficient and necessary condition for the weak Lie-invariance to be equivalent to the strong Lie-invariance of $\psi$ along $\xi$}.

Because the equivalence of strong and weak Lie-invariance is subject to condition (\ref{COND}), and since this condition depends on the vector $P_{\mu}\!-\!V_{\mu}$, it would be interesting now to discuss \eqref{COND} more in detail.

In a variational approach, the energy of a free spinor field in flat space-time, in Cartesian coordinates, is the conserved current of the symmetry given by time translations. Within the Lie analysis, time-independence is implemented by the Killing vector field
\begin{eqnarray}
&\xi\!=\!\left(\begin{array}{c}
\!1\!\\
\!0\!\\
\!0\!\\
\!0\!
\end{array}\right)
\end{eqnarray}
as well known. Because this means $\partial \xi\!=\!0$ we have that \eqref{COND} is reduced to $\xi^{\mu}(P_{\mu}\!-\!V_{\mu})\!=\!0$ identically. Hence 
\begin{eqnarray}
&T_{\rho\sigma}\xi^{\rho}\xi^{\sigma}\!=\!\phi^{2}\xi_{\rho}s_{\kappa}u_{\alpha}u^{\mu}\nabla_{\mu}\xi_{\nu}\varepsilon^{\nu\rho\kappa\alpha}\!=\!0
\end{eqnarray}
because of weak Lie-invariance and again the fact that $\partial \xi\!=\!0$ identically. Therefore, time-independence gives not only energy conservation but also energy vanishing, when condition (\ref{COND}) happens to be verified.

In this case, while strong Lie-invariance is mathematically possible, it may be dynamically too restrictive.
\section{Spherical Symmetry}\label{SecSpher}
In the previous sections we have found the condition under which weak and strong Lie-invariance are equivalent to each other and discussed that this condition might be too restrictive on dynamical grounds. This means that in very general situations there is no equivalence between weak and strong Lie-invariance, so that only the weak Lie-invariance should be required to implement a symmetry on a given system, if we want to embrace the less restrictive case.

However, even the case of weak Lie-invariance may lead to contradictions in some cases. One of these cases is that of spherical symmetry. In this situation in fact, the Killing vector fields, in spherical coordinates, are given by
\begin{eqnarray}
&\xi_{0}\!=\!\left(\begin{array}{c}
\!1\!\\
\!0\!\\
\!0\!\\
\!0\!
\end{array}\right)\ \ \ \ \ \ \ \ \xi_{1}\!=\!\left(\begin{array}{c}
\!0\!\\
\!0\!\\
\!-\cos{\varphi}\!\\
\!\sin{\varphi}\cot{\theta}\!
\end{array}\right)\ \ \ \ \ \ \ \ \xi_{2}\!=\!\left(\begin{array}{c}
\!0\!\\
\!0\!\\
\!\sin{\varphi}\!\\
\!\cos{\varphi}\cot{\theta}\!
\end{array}\right)\ \ \ \ \ \ \ \ \xi_{3}\!=\!\left(\begin{array}{c}
\!0\!\\
\!0\!\\
\!0\!\\
\!1\!
\end{array}\right)
\end{eqnarray}
and therefore the metric is
\begin{eqnarray}
&g_{tt}=e^{2A}\ \ \ \ \ \ \ \ \ \ g_{rt}\!=\!e^{(A+B)}\sinh{\eta}
\ \ \ \ \ \ \ \ \ \ g_{rr}=-e^{2B}\ \ \ \ \ \ \ \ \ \
&g_{\theta\theta}=-e^{2C}\ \ \ \ \ \ \ \ \ \ g_{\varphi\varphi}=-e^{2C}|\sin{\theta}|^{2}
\end{eqnarray}
where $A=A(r)$, $B=B(r)$, $C=C(r)$ and $\eta=\eta(r)$ are generic functions of the radial coordinate. Notice that the case $\eta\!\rightarrow\!0$ is the one for which the metric becomes diagonal, implementing the passage from stationary to static case, while the restriction $C\!\rightarrow\!\ln{r}$ makes the metric reduce to Schwarzschild-like space-times. The weak Lie-invariance implies that the scalars $\phi$ and $\beta$ can only be functions of the radial coordinate and that the vectors $u^{\alpha}$ and $s^{\alpha}$ can only have the time and radial components and that these are functions of the radial coordinate, or explicitly
\begin{eqnarray}
&u_{t}=e^{A}\cosh{(\alpha\!+\!\eta)}\ \ \ \ \ \ \ \ \ \ \ \ \ \ \ \ 
u_{r}=-e^{B}\sinh{\alpha}\label{vec}\\
&s_{t}=e^{A}\sinh{(\alpha\!+\!\eta)}\ \ \ \ \ \ \ \ \ \ \ \ \ \ \ \ 
s_{r}=-e^{B}\cosh{\alpha}\label{ax-vec}
\end{eqnarray}
where $\alpha\!=\!\alpha(r)$ and in which the constraints \eqref{norm} have also been taken into account. We notice that for $\eta\!=\!0$ the case $\alpha\!\rightarrow\!0$ is the one for which we find ourselves in the spinor rest frame and with spin aligned along the radial axis.

The invariance under parity is implemented by the conditions $s_{\nu}'\!=\!s_{\nu}$ and $u_{\nu}'\!=\!u_{\nu}$ for the transformation $\theta'\!=\!\pi-\theta$, which in particular implies $\partial\theta'/\partial\theta\!=\!-1$ and a negative Jacobian: because there is no $\theta$ component and all those that exist do not depend on $\theta$, parity acts only via the Jacobian, as $s_{t}'\!=\!-s_{t}$ and $s_{r}'\!=\!-s_{r}$, and that is $s_{\nu}'\!=\!-s_{\nu}$, with $u_{\nu}$ left unchanged. So, the invariance under parity would imply $s_{\nu}\!=\!s_{\nu}'\!=\!-s_{\nu}$, and therefore $s_{\nu}\!=\!0$ identically: thus we have that $s^{a}s_{a}\!=\!0$ and this is incompatible with the fact that $s^{a}s_{a}\!=\!-1$, generating a contradiction. Because such a condition is a scalar condition, it does not depend on the spherical coordinates, and the result is fully general.
\section{Conclusion}
In this work, we have highlighted the role of the polar decomposition of spinors in discussing their Lie derivative along Killing vector fields, clarifying the relationship between the Lie derivative of the spinor and that of its observables. After introducing the notions of strong Lie invariance (that is the vanishing of the Lie derivative of the spinor) and weak Lie invariance (that is the vanishing of the Lie derivative of all the spinor bi-linears), we have singled out the condition under which the two notions are equivalent.

As an example, we have proved that no spinor field undergoing spherical symmetry via Lie derivative can be defined. Because the proof has been obtained in terms of  scalar conditions employing polar variables at the kinematic level, the result does not depend on the chosen coordinates or tetrads nor any interaction.

\

\textbf{Funding information and additional acknowledgements}. This work has been carried out in the framework of activities of the INFN Research Project QGSKY. The work of L.F. has been funded by Next Generation EU through the project ``Geometrical and Topological effects on Quantum Matter (GeTOnQuaM)''.

\

\textbf{Data Availability Statement}. The manuscript has associated data in a repository.

\

\textbf{Conflict of interest}. There is no conflict of interest.

\end{document}